\documentclass[6pt, preprint2]{aastex}
\usepackage[]{natbib}

\begin{document}

\title{Six Peaks Visible in the Redshift Distribution of 46,400 SDSS Quasars Agree with the Preferred Redshifts Predicted by the Decreasing Intrinsic Redshift Model}

\author{M.B. Bell\altaffilmark{1} and D. McDiarmid\altaffilmark{1}}
\altaffiltext{1}{Herzberg Institute of Astrophysics,
National Research Council of Canada, 100 Sussex Drive, Ottawa,
ON, Canada K1A 0R6;
morley.bell@nrc.ca}

\begin{abstract}

The redshift distribution of all 46,400 quasars in the Sloan Digital Sky Survey (SDSS) Quasar Catalog III, Third Data Release, is examined. Six Peaks that fall within the redshift window below z = 4, are visible. Their positions agree with the preferred redshift values predicted by the decreasing intrinsic redshift (DIR) model, \em even though this model was derived using completely independent evidence. \em A power spectrum analysis of the full dataset confirms the presence of a single, significant power peak at the expected redshift period. Power peaks with the predicted period are also obtained when the upper and lower halves of the redshift distribution are examined separately. The periodicity detected is in linear z, as opposed to log(1+z). Because the peaks in the SDSS quasar redshift distribution agree well with the preferred redshifts predicted by the intrinsic redshift relation, we conclude that this relation, and the peaks in the redshift distribution, likely both have the same origin, and this may be \em intrinsic redshifts \em or \em a common selection effect. \em However, because of the way the intrinsic redshift relation was determined it seems unlikely that one selection effect could have been responsible for both.

\end{abstract}


\section{Introduction.}

Peaks in the distribution of quasar redshifts have been claimed as evidence for the existence of preferred redshifts, and a periodicity in log(1+z) was suggested after only a few hundred redshifts were known \citep{kar71,kar77,bur01}. Since this claimed periodicity came from the early redshift samples themselves, until more complete samples became available it was not possible to test it. Recently, using a SDSS quasar redshift sample containing approximately 5,000 high-redshift quasars, it was demonstrated that there is little evidence that the previously claimed log(1+z) periodicity fits the peaks in the SDSS redshift distribution above z = 2 \citep{bel04}. However, it was demonstrated previously that the redshift peaks found in the early data can be fitted reasonably well to a different intrinsic redshift relation that is quasi-periodic in linear redshift, as opposed to log(1+z) \citep{bel02a,bel02c,bel03a}. The intrinsic, or preferred, redshift components predicted in this model, z$_{iQ}$, are given by the relation

z$_{iQ}$ = z$_{f}$[$N$ - $M_{N}$] ---------------- (1)
 
where $N$ is an integer, $M_{N}$ is a function of a second quantum number $n$, defined previously \citep{bel02c,bel03a}, and z$_{f} = 0.62\pm0.01$ is the intrinsic redshift constant. Unlike the log(1+z) relation, this equation did not come from an analysis of peaks in a redshift distribution. It was determined instead, after all Doppler components were estimated and removed from the redshifts of the 14 QSOs near NGC 1068 \citep{bur99,bel02a,bel02b,bel02c}. The intrinsic redshift components that led to eqn 1 were obtained by fitting an ejection model to the sources. This model used both the measured redshifts, and the distances of the sources from NGC 1068 to estimate ejection velocities before calculating the intrinsic redshift component.

 All redshift components predicted by eqn 1 are listed in Table 1, for values less than z = 4.5, and have been available in the literature for over three years. As can be seen from Table 1, eqn 1 becomes periodic in $\Delta$z = 0.62 for N $> 4$. Below N = 5, each N-group contains additional redshift components that in all cases fall below the z = 0.62N harmonic. Searching a quasar redshift distribution as large as that provided by the SDSS Third Data Release, for the preferred redshift values predicted by eqn 1, represents \em a completely independent test of this equation. \em
%A reasonably complete statistical sample of quasar redshifts such as these SDSS %quasars can be used to examine the long-standing controversy surrounding the %nature of quasar redshifts.
%Furthermore, here we propose to use all redshifts, unlike some previous %analyses which used only selected redshift samples \citep{arp05}. Our arguments %for this are as follows. We have some concern that sometimes after the value of %a parameter has been found, selecting data sub-samples can result in a fishing %expedition to find sub-samples that, by chance, strengthen the earlier finding, %but whose results then will have little significance. If a feature found in an %early data sample is real, it must continue to be present as more sources are %added to the sample, unless the sampling criteria change.

Evidence was presented recently showing that peaks in the distribution of SDSS quasar redshifts near z = 3.1 and 3.7 had extended high-redshift wings that could be attributed to the presence of a small cosmological redshift component z$_{c} < 0.066$ \citep{bel04}. Here the redshift distribution for 46,400 redshifts in the full SDSS Catalog III sample \citep{sch05} is examined to see if there is evidence for a redshift periodicity that would be expected to result from the quasi-periodic preferred redshift values predicted by eqn 1. First, we carry out a power spectrum analysis on the entire SDSS redshift distribution, containing over 46,000 quasar redshifts. We then visually compare the SDSS redshift distribution to the quasi-periodic values predicted by equation 1 to demonstrate that the peaks found are not only in good agreement with the fundamental periodicity, but also with the sub-components predicted by eqn 1. We also obtain an independent estimate of the errors, and examine selection effects that might play a role in producing peaks in the distribution. Finally, we carry out spectral analyses on both the low and high-redshift halves of the data.

%Although currently thought to be simply a coincidence, as noted previously %\citep{bel04} the intrinsic redshift constant, z$_{f}$, is also equal, within %the uncertainty, to the so-called \em golden ratio, \em or, by convention, its %reciprocal, $1/\phi = 0.618$ \citep{liv02}.

\begin{figure}
\hspace{-1.0cm}
\vspace{-2.0cm}
\epsscale{1.0}
\plotone{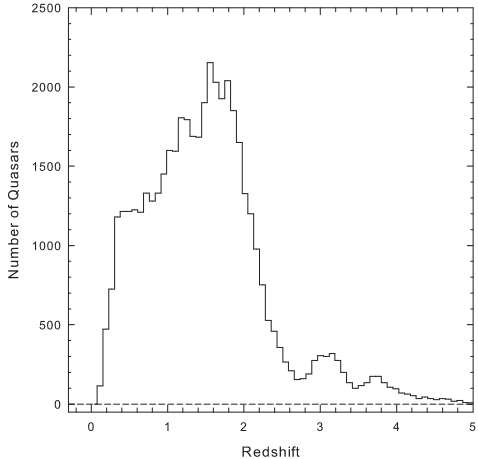}
%\plotone{nvszv2.eps}
\caption{\scriptsize{Distribution of 46,400 SDSS quasars with redshifts below z = 5 \citep{sch05}. \label{fig1}}}
\end{figure}

\begin{figure}
\hspace{-1.0cm}
\vspace{-1.5cm}
\epsscale{1.0}
\plotone{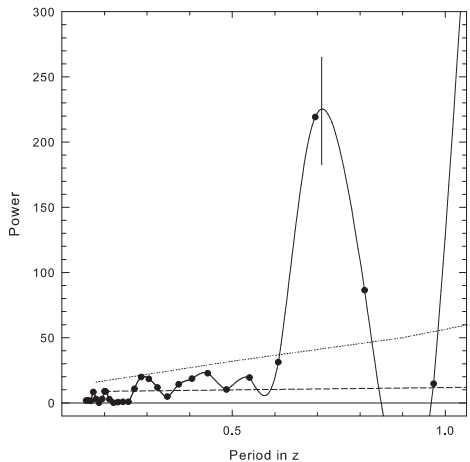}
%\plotone{fftallpowervsperiodv2.eps}
\caption{\scriptsize{(solid curve)Spectral power obtained for over 46,000 quasar redshifts plotted vs redshift period. The the vertical bar and dotted line represent the $1 \sigma$ error taken from \citet{tan05}. The dashed line represents the statistical uncertainty obtained for 46000 randomly generated redshifts. Negative values are introduced by the spline fitting. \label{fig2}}}
\end{figure}

\begin{figure}
\hspace{-1.0cm}
\vspace{-1.5cm}
\epsscale{1.0}
\plotone{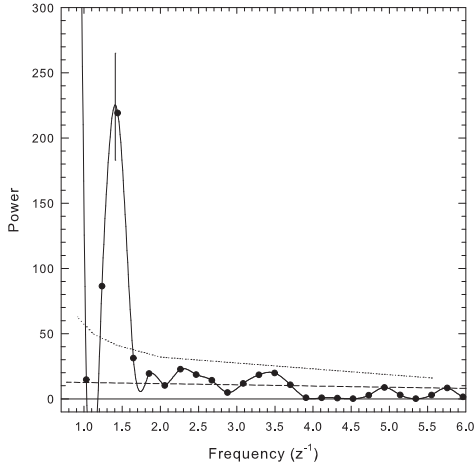}
%\plotone{fftbasepowervsfreqv2.eps}
\caption{\scriptsize{Spectral power distribution obtained for over 46,000 quasar redshifts plotted vs frequency (z$^{-1}$). See text for a description of the vertical bar. Dotted and dashed lines are as in Fig 2. \label{fig3}}}
\end{figure}

\begin{figure}
\hspace{-1.0cm}
\vspace{-1.5cm}
\epsscale{1.0}
\plotone{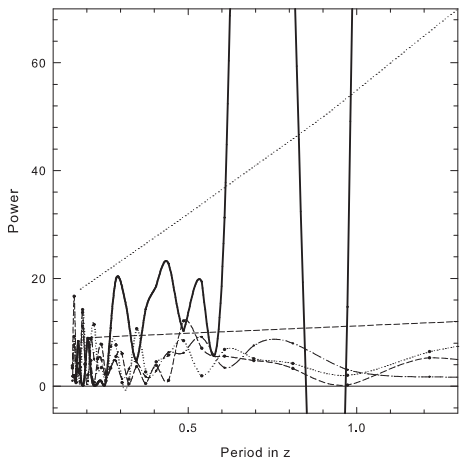}
%\plotone{sigma3000.eps}
\caption{\scriptsize{(solid) FFT of the distribution 46,400 SDSS redshifts, (dashed, dotted and dashed-dot curves) FFT's of 46000 randomly generated redshifts. (dashed and dotted lines) uncertainties as in Figs 2 and 3. \label{fig4}}}
\end{figure}

\section{Observations}

If, as is assumed in the DIR model, quasars are ejected from active galaxies, and these galaxies are distributed uniformly in space, the same should be true for quasars, assuming that the ejection process is similar at all epochs. However, since in this model they are sub-luminous by several magnitudes when first born, distant, young quasars (QSOs) will not be detectable with current sensitivities. On the other hand, those with lower intrinsic redshifts, which are more luminous than those with high intrinsic redshifts \citep{bel02a,bel02c}, will be detectable to greater cosmological distances. It is also worth noting again that in the DIR model the intrinsic component present in quasar redshifts is superimposed on top of a second redshift component that has been found, at least in the local Universe, to be indistinguishable from that of a Hubble flow with H$_{o}$ = 58 km s$^{-1}$ Mpc$^{-1}$ \citep[and references therein]{bel04a}.

In Fig 1, the histogram shows the distribution of 46,400 quasar redshifts from the SDSS Third Data Release \citep{sch05}. The sources have been binned into redshift intervals with $\Delta$z = 0.076. The gross features in the redshift distribution, with a large, broad component below z = 2.2 and a low pedestal above z = 2.4, have resulted from data selection effects. In the low-redshift source selection sample (z $< 2.2$) an $i$ magnitude limit of 19.1 was imposed for candidates whose colors indicated a probable redshift of less than $\sim3$ (selected from the $ugri$ color cube); in the high redshift sample candidates (selected from the $griz$ color cube) are accepted if $i < 20.2$. A detailed description of the quasar selection process and possible biases can be found in \citet{ric02}. As can be seen from \citet[figs 2 and 4]{sch05}, a) there are many more sources in the low redshift sample than in the high-redshift sample and, b) the upper cut-off of the low-redshift sample is abrupt. The broad structure introduced by the low-redshift source selection process is expected to produce mainly long-period components in the power spectrum.
%it is possible that the rather abrupt cut-off in the distribution near z = 2 %may introduce some spectral components with shorter redshift periods. This is %also examined below.

\subsection{Power Spectrum Analysis}

The power spectrum analysis of the entire data sample was carried out by taking a fast Fourier transform of the histogram in Fig 1 for the 64 channels between z = 0 and z = 4.86.
The power obtained is plotted vs redshift period as the solid curve in Fig 2, and versus frequency (z$^{-1}$) in Fig 3. In each case the negative values have been introduced by the spline fitting. The vertical solid bar indicates the one standard deviation uncertainty estimated by \citet{tan05} who carried out a similar power spectrum analysis on the same data sample. It was taken from their Fig 9. The dotted curve represents this approximate 1 std. dev. uncertainty plotted as a function of period and frequency respectively. Estimating the uncertainty in power spectra can be a difficult process, and even when great care is taken the result needs to be treated with caution. As can be seen from \citet[Fig 1]{tan05}the size of 1 std. dev. increases with power and caution must be taken if one attempts to use it to obtain a measure of the significance of the power peak. We independently determined the uncertainty that could be expected to be introduced due to statistical fluctuations in the data when there are no selection effects present. To do this we generated three random sets of 46,000 redshifts, uniformly distributed between z = 0 and 4.86. The redshifts in each set were binned into 64 bins with a width of $\Delta$z = 0.076, exactly as was done for the real data. A power spectrum was then obtained for each of the three data sets and the results are plotted in Fig 4 on an expanded power scale where the peak level (dashed line) can be compared to the 1 $\sigma$ value obtained by \citet{tan05}(dotted line). Our error represents the peak error that is produced by random, uniformly distributed, fluctuations in a sample of 46,000 redshifts, plotted as a function of the harmonic number.
%This error does not include spectral components that will be introduced by a) %selection effects, b) real redshift fluctuations due to preferred values, or c) %leakage components that are introduced by edge effects in the Fourier %processing. Because of 
This error level is significantly lower than the error set by \citet{tan05}, and is plotted as a dashed line in Figs 2, 3 and 4.
%From the description of their error calculation it is not clear if it includes %a contribution from real components that may be present in the redshift %distribution, or if it just includes those produced by the spectral analysis %when such components are present. 
%As noted above, the error they show (see their Fig 1) \em increases as the %signal power increases. \em This would seem to imply that the S/N ratio would %remain constant regardless of how strong the signal became.

\citet{tan05} found, as found here, that there is a significant periodicity with period near 0.7 in redshift in the full sample containing over 46,000 redshifts, and regardless of which error estimate is used, the question that now needs to be answered is whether this power peak is due to the presence of preferred redshifts or to selection effects.

The strong power level above $\Delta$z $\sim$1 arises almost entirely from the shape of the distribution created by the low-redshift data selection effect discussed above.

%For periods shorter than z = 1 there is clear evidence for only one strong %power peak in the spectrum with period near 0.7.  It represents at least 6 std. %dev. based on the Tang and Zhang uncertainties. This is closer to 25 using our %calculations. This peak may also be due to a selection effect; however, this %period is close to that predicted by eqn 1, which means it might also have been %produced by the preferred redshifts predicted by the DIR model.

%It is important to note that this analysis was carried out using the entire %redshift distribution given by the histogram in Fig 1, prior to the removal of %the smooth base component of the low-redshift sample. The vertical bar in Fig 3 %indicates the location of the redshift period estimated from Fig 2, and the two %clearly show excellent agreement.

%Although it is expected that the broad shape of the low-redshift sample (below %z = 2.4) (approximated by the smooth curve in Fig 1) will contribute mainly %long-period Fourier components to the power spectrum, it is also of interest to %confirm this. The dashed line in Fig 3 represents the spectral power produced %by the broad, smooth, base component alone, whose power, as predicted, is %mainly concentrated at the long period end of the distribution. To maintain %adequate resolution in this analysis we again used 64 points, obtained by %padding the 32 values above z = 2.43 with zeros. This dashed curve thus gives %us an indication of how the data selection effects discussed above contribute %to the Fourier components present in the distribution.

\begin{figure}
\hspace{-1.0cm}
\vspace{-2.0cm}
\epsscale{1.0}
\plotone{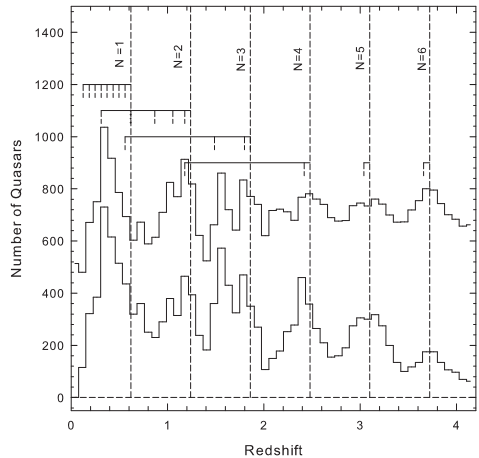}
%\plotone{nvsz2345v2.eps}
\caption{\scriptsize{(lower histogram) Distribution of SDSS quasar redshifts after removal of the curved baseline in Fig 6. (upper histogram) Distribution of quasar redshifts obtained after taking the Fourier transform, setting the first four non-DC Fourier components to zero, and then taking the inverse Fourier transform. The vertical dashed lines indicate the positions of the highest redshift in each N-group from eqn 1. The vertical short dashed lines indicate redshifts from eqn 1 that fall below the maximum in each N-group. \label{fig5}}}
\end{figure}

\begin{figure}
\hspace{-1.0cm}
\vspace{-2.0cm}
\epsscale{1.0}
\plotone{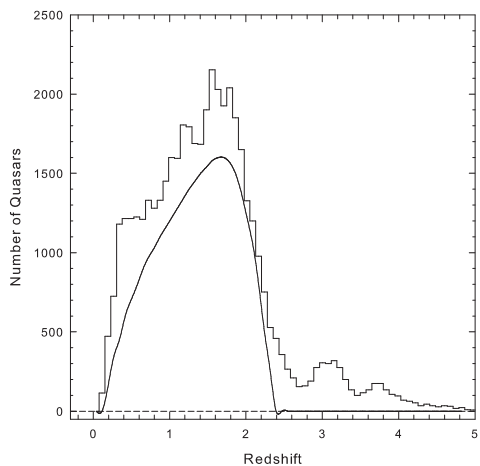}
%\plotone{nvsz7.eps}
\caption{\scriptsize{(upper histogram)Distribution of 46,400 SDSS quasars with redshifts below z = 5 \citep{sch05}. See text for a description of the curved baseline below z = 2.4. \label{fig6}}}
\end{figure}

%\begin{figure}
%\hspace{-1.0cm}
%\vspace{-1.5cm}
%\epsscale{1.0}
%\plotone{diffpowervsfreq.eps}
%\caption{\scriptsize{Spectral power distribution obtained for the difference %histogram in Fig 2, plotted vs frequency (z$^{-1}$). The vertical bar indicates %the frequency estimated from Fig 2. \label{fig6}}}
%\end{figure}

\section{Comparison between the data peaks and the predicted preferred redshifts}

Above, we found that there was a significant power peak in the redshift distribution with a period near $\Delta$z = 0.7. Equation 1 predicts preferred redshifts that are quasi-periodic with a slightly shorter period of $\Delta$z = 0.62. Here we examine the redshift distribution together with the redshifts predicted by eqn 1 to see if this difference can be explained.  Although several peaks are already visible in the raw redshift distribution in Fig 1, before undertaking this comparison it is first helpful to remove the broad, smooth baseline component below z = 2.4, on top of which the peaks are superimposed. We did this using two different techniques, and the results are shown in Fig 5. First we subtracted a smooth, low-redshift, baseline curve. The curve used is shown by the smooth curve in Fig 6 and its shape, with a relatively sharp upper cut-off, is based on the shape of the number vs magnitude distribution produced by the low-redshift sub-sample \citep[see Fig 4]{sch05}, and the fact that there is a relation between redshift and magnitude. In the DIR model this broad, featureless curve is attributed to sources whose cosmological expansion, or Doppler ejection, redshift components are comparable to, or larger than, the spacing between the predicted intrinsic values. The resulting redshift distribution curve, after subtraction of the smooth baseline curve, is shown by the lower curve in Fig 5. However, because the shape of this smooth baseline curve is somewhat arbitrary it is conceivable that its removal may have introduced some spurious features. This is especially true near z = 2.5 where the curve ends abruptly. Therefore we also used a second, more objective, method to remove the broad, low-redshift selection effect. In this method, after obtaining the Fourier components for the entire sample, the first four non-DC Fourier components were then set to zero. This effectively filters out all long-period fluctuations. The inverse Fourier transform was then obtained and the result is plotted in the upper curve in Fig 5. Both techniques gave similar results and 6 peaks between z = 0 and z = 4 are clearly visible in each curve. In Fig 5, the vertical dashed lines indicate the positions of the highest redshift in each N-group as defined by equation 1. The shorter dashed lines indicate the positions of the lower redshift sub-components in each N-group. Since these components are predominantly present in only the first three N-groups, and all lie at lower redshifts, this will introduce an effective stretching out of the peak separation at the low redshift end of the distribution. This effect is most clearly visible in the first N-group where the peak is centered near the mean N = 1 group redshift of z = 0.31.

In addition to this stretching at the low-redshift end of the distribution, if there is a small cosmological component present as argued previously \citep{bel04}, this will also tend to stretch out the high redshift end of the distribution. Using the relation (1+z) = (1+z$_{c}$)(1+z$_{i}$), where z$_{c}$ is the cosmological component and z$_{i}$ is the intrinsic one, for a mean cosmological component of z$_{c}$ = 0.02 the intrinsic redshift peak at z$_{i}$ = 3.72 would be stretched to z = 3.81. It is therefore possible to obtain a rough estimate of how significant the total stretching effect will be when the redshifts predicted from eqn 1, together with the stretching at both the low and high ends of the distribution are taken into account. Since there are 5 N-cycles between z = 0.31 and z = 3.81, a period of $\Delta$z = 0.70 is estimated from Fig 5, for a mean cosmological redshift of z$_{c}$ = 0.02. This period agrees well with the power peak obtained. Close examination of Fig 5 reveals that not only are there peaks associated with all harmonics of 0.62 below z = 4, there is also reasonable agreement below z = 2 with the predicted redshift sub-components. For example, the peaks at .31 and 1.1 coincide with regions where there is a high density of preferred redshift components. Also, the double peaks at z = 1.55 and 1.85 agree well with the predicted preferred redshifts of 1.488 and 1.798, if a small cosmological component is present. 

\begin{figure}
\hspace{-1.0cm}
\vspace{-1.5cm}
\epsscale{1.0}
\plotone{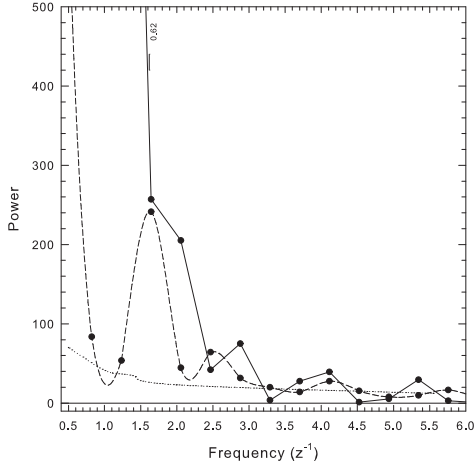}
%\plotone{fftch1to32vsfreqnopadv2.eps}
\caption{\scriptsize{(solid) FFT of bottom 32 bins of the redshift range, (dashed) FFT of upper 32 bins. \label{fig7}}}
\end{figure}

\begin{figure}
\hspace{-1.0cm}
\vspace{-1.5cm}
\epsscale{1.0}
\plotone{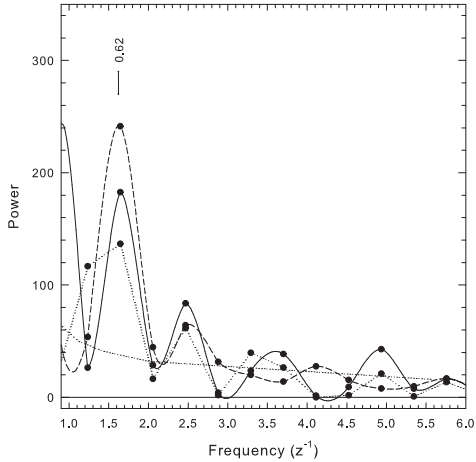}
%\plotone{fftch1to32lessbasevsfreqnopadv2.eps}
\caption{\scriptsize{Power vs frequency. (dashed line) upper half of redshifts. ((solid line) lower half of redshifts after smooth baseline removed. (dotted line) lower half of redshifts with first 4 harmonics removed.  \label{fig8}}}
\end{figure}

%\begin{figure}
%\hspace{-1.0cm}
%\vspace{-1.5cm}
%\epsscale{1.0}
%\plotone{f4.eps}
%\plotone{fftch1to32vsfreqoncepadv2.eps}
%\caption{\scriptsize{(solid curve)Spectral power distribution obtained for %quasar redshifts below z = 2.43 plotted vs frequency (z$^{-1}$). (dashed curve) %Spectral power obtained for the upper half of the redshift distribution $(2.4 < %4.8)$.  Vertical bar and dotted line are as in Fig 2. \label{fig8}}}
%\end{figure}

%We also investigated how the shape of the broad, low-redshift distribution %curve might affect the resulting difference histogram and found that only near %the cut-off (near z = 2) can its subtraction significantly affect the shape of %the resulting histogram in Fig 2. Depending on how steep the cut-off is, the %depth of the valley near z = 2, and the height of the peak near z = 2.4, can %both be altered. The cut-off slope used was chosen so that this peak and valley %looked similar to others in the histogram in Fig 2. Note that the location of %these features is not affected, only their levels.

\section{Spectral Analysis of the Upper and Lower Halves of the Redshift Distribution}

Because the detection efficiency of the SDSS is lower near z = 2.7 and z = 3.5, this could have contributed to the depth of the valleys seen in the redshift distribution at these redshifts \citep{sch05}. It is therefore of interest to see if there is a contribution to the observed power peak that comes solely from the lower half of the redshift distribution (z $< 2.4$), keeping in mind, that, because the density of lines predicted by equation 1 is higher in this redshift range, preferred redshifts will be more easily smeared out if small cosmological redshift components are present.

Fig 7 shows power spectra obtained by splitting the redshift range into two halves. The spectrum of the lower half of the redshift range (32 bins below z = 2.4) is shown by the solid line and that of the upper half by the dashed line. Because only half of the redshift range has been covered here, the spectral resolution is reduced. Because of this the strong low-frequency components present in the low-redshift half of the data make it impossible to resolve completely the peak near z$^{-1}$ = 1.5. Although there is evidence for an unresolved feature in the solid curve in Fig 7, it will be necessary to remove the low-frequency components before a convincing power peak can be resolved. 
%we took two different approaches. We first obtained the power spectrum over 64 %bins by padding the upper 32 bins with zeros. The results are plotted in Fig 8 %where a double-peaked, but clearly resolved, feature is now apparent near z$^{-%1}$ = 1.5 in the solid curve. Second,
To do this we removed the low-frequency Fourier components exactly as was done for the curves in Fig 5, and recalculated the FFT. In Fig 8 the results are again compared to the power spectrum obtained for the upper half of the redshifts. Here the solid curve was obtained using the lower curve in Fig 5, and the dotted curve corresponds to the upper curve. There is now clear evidence for a power peak near a frequency of 1.6 in the lower half of the redshift data in Fig 8. Since \citet{tan05} made no effort to remove the overwhelming effects of the strong low-frequency components when they examined the lower half of the redshift data, they would not have been able to detect this feature. But this should not be too surprising since \em these authors also failed to detect a significant power peak near $\Delta$z = 0.62 in the high redshift sample, even though one is clearly visible. \em The reason for this is easily seen. For some reason these authors extended the redshift range down to z = 2, which covers a portion of the much more highly populated low-redshift source sample. By so doing they included the large transition step between the two redshift samples that is produced solely by this selection effect. This would clearly wreak havoc with the transform. The resulting strong, longer-period power peak near 0.75 introduced by the portion of the redshift distribution between z = 2 and z = 2.4 (see their Fig 11 (d)) has simply overwhelmed the peak at $\Delta$z = 0.62, and prevented its detection. We demonstrate this in Fig 9, where the solid line represents the spectral power obtained for 32 bins above z = 2. The dashed line represents the power spectrum obtained for 32 bins above z = 2.4. There is no evidence for a power peak at $\Delta$z = 0.62 in the former, but one is clearly visible in the latter.

%This portion of Tang and Zhang's analysis may bring into question their entire %paper.

The strong peak in the upper half of the data shown as a dashed line in Figs 7 and 8 has a period P of $\Delta$z = 0.62. That this period, over a redshift range where few sub-components are present, is exactly equal to the harmonic constant z$_{f}$, is further evidence that the valleys near z = 2.7 and 3.5 may \em not \em be caused entirely by the suggested selection effect, and this is discussed further in the next section.

It is also worth noting that when periodicities are sought in the low-redshift half of the data, this transition step between z = 2 and z = 2.4 also plays a role in preventing detection of the period being sought. When a power spectrum is obtained for the redshift range from z = 0 to z = 2, a significant power peak near the expected frequency is obtained immediately, without any further need to remove low-frequency components.

The failure by \citet{tan05} to detect a strong power peak at $\Delta$z = 0.62 in the upper half of the redshift range is also a strong argument that dividing the data into smaller sub-samples can give misleading results if not done carefully.

\begin{figure}
\hspace{-1.0cm}
\vspace{-1.5cm}
\epsscale{1.0}
\plotone{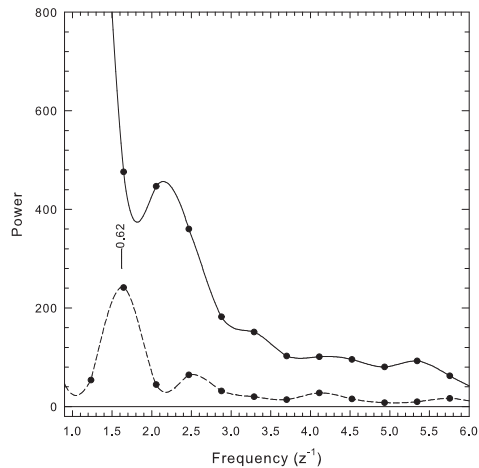}
%\plotone{fft32chabove2vsfreqv2.eps}
\caption{\scriptsize{Spectral Power vs frequency. (solid curve) 32 bins above z = 2.0. (dashed curve) 32 bins above z = 2.4.  \label{fig9}}}
\end{figure}

\begin{figure}
\hspace{-1.0cm}
\vspace{-1.5cm}
\epsscale{1.0}
\plotone{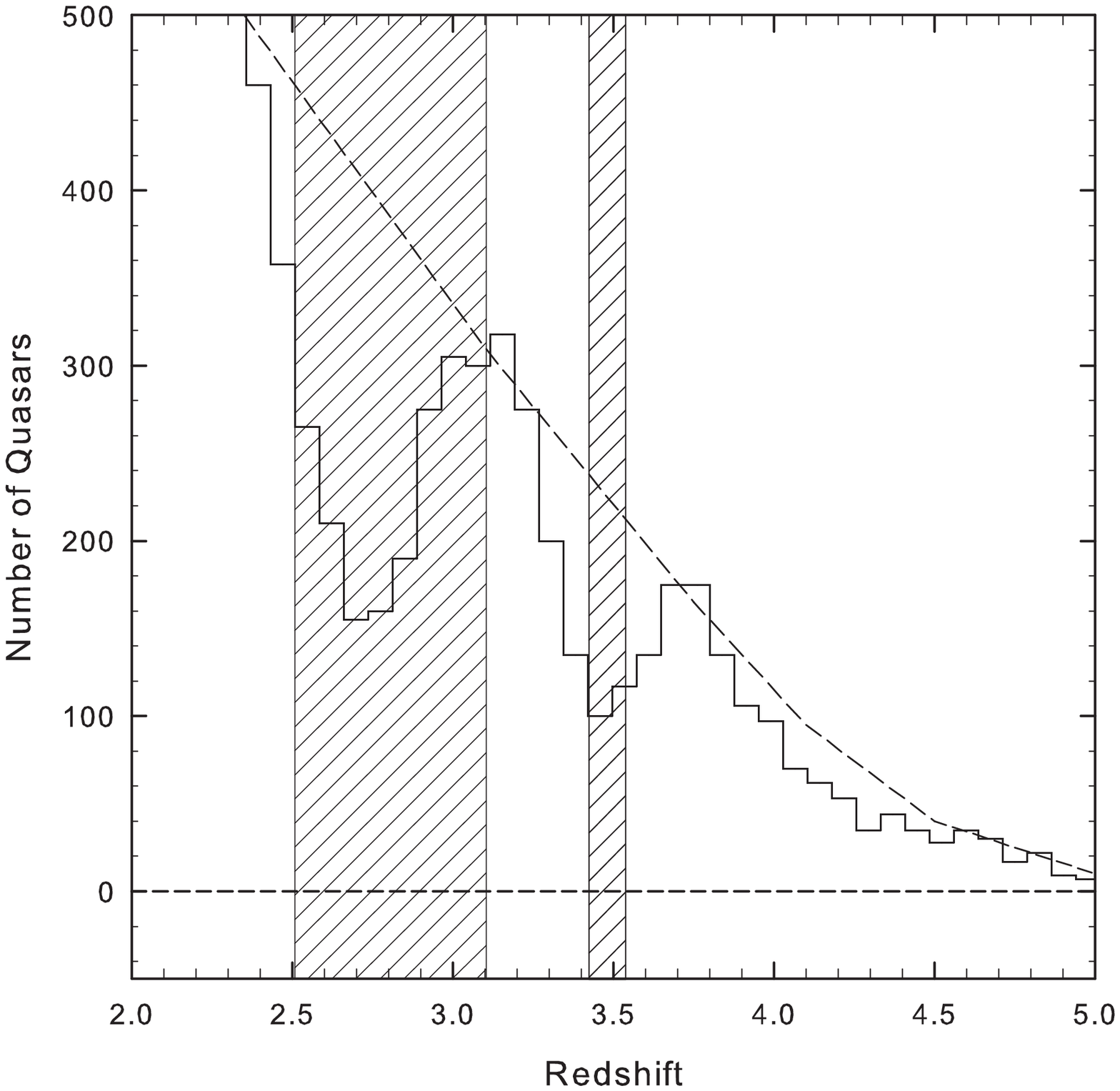}
%\plotone{nvszv2high2.eps}
\caption{\scriptsize{Crosshatcged areas show the regions where valleys are expected to appear from \citet[Fig 10]{ric02}. See text for further comments.  \label{fig10}}}
\end{figure}

\section{Selection effects in the data}

Just because there appears to be good agreement between the peaks in the redshift distribution and the preferred redshifts predicted by eqn 1, it does not necessarily mean that intrinsic redshifts are real. As mentioned above, there are selection effects that have been claimed to explain the valleys in the distribution near z = 2.7 and 3.5, and there could conceivably also be other selection effects responsible for the other valleys in the distribution.
%However, because of the good agreement we find between the peaks and the %predicted, preferred redshifts, it seems likely that if selection effects are %the cause, \em they must be able to explain both the peaks in the distribution, %as well as eqn 1. \em
Because quasar colors are similar to those of stars at redshifts near z = 2.7 and z = 3.5, finding all the quasars at these redshifts would be a formidable task, since it would require measuring the redshifts of all the stars as well. The strategy used to avoid this could have resulted in the valleys at these redshifts. However, we also point out that eqn 1 predicts that no quasars will have these preferred redshift values, except for a few that have been shifted there because they also contain a large cosmological component or a large ejection velocity component, or both. In this case the suggested selection effect could not play a significant role. If there are few sources there, finding only a small fraction of them would not change the result, and even if all sources were examined, no quasars would be found. The only way to prove that the suggested selection effect is causing the dips is then to measure every object with these colors. Until this happens, the correct explanation of these dips will remain unknown. 

However, there may be other evidence to indicate that the suggested selection effect is not the only explanation for the high-redshift valleys. In Fig 10 the high-redshift portion of the distribution has been replotted. The valleys near z $\sim2.7$ and z $\sim3.5$ have similar widths, with depths that are approximately fifty percent of the peaks estimated by the dashed line. However, Fig 10 of \citet{ric02} shows that the dip at z $\sim3.5$ should be much narrower and shallower than the one at z $\sim2.7$. The crosshatched areas in Fig 10 show the regions where the valleys are expected and there is poor width agreement for the valley at z $\sim3.5$. Also, from \citet[Fig 10]{ric02} this dip appears to be much deeper, relative to the one at z = 2.7, than would be expected if it were due to the suggested selection effect.

It has also been suggested that optical observing effects, such as strong emission lines redshifting through the observing window, and the availability of search lines at certain redshifts, etc, could result in sources being more easily detected at certain redshifts \citep[and references therein]{bas05}. Since the quasars in the SDSS sample are identified optically, the redshift distribution could conceivably have been influenced by these kinds of optical selection effects if they are significant.
%However, to the best of our knowledge there have been no reasons suggested to %explain the valleys in the distribution below z = 2.
In Fig 11 the relation between redshift and the observed wavelengths of the strong emission lines is shown. Here the horizontal dashed lines represent the locations of the gaps between the SDSS filters \citep[see Fig 4]{ric02}. It is easily seen at which redshifts the strong lines are located inside each filter passband. Included in the figure are vertical dotted lines representing the locations of the peaks found in the redshift distribution. The locations of the valleys are indicated by the vertical solid lines. There are several observations that can be made concerning this plot. First, the peak at z $\sim0.3$ occurs at a redshift that places the strong Hydrogen Balmer lines (H$_{\alpha}$ and H$_{\beta}$) in the gaps between filters. This is contrary to what would be expected if these lines were influencing the source selection at this redshift. Second, the valley near z = 0.8 occurs when the MgII and H$_{\beta}$ lines are in the g and i filter passbands, respectively, again contrary to what would be expected if these line were influencing the source selection. There are other examples, some that agree and some that disagree, but there appears to be little correlation between the presence of these strong lines in the filter passbands, and the presence of peaks in the redshift distribution.
%There also appears to be no correlation between the valleys in the SDSS %distribution and the passing of strong emission lines through the less %sensitive wavelengths between the filter passbands. 
 
%Concerning source samples that have been obtained from radio surveys, it has %been argued that it is the redshift measurement that ultimately determines the %shape of the distribution, and not the radio finding survey \citep{bas05}. %However, there is no evidence that many of the sources found in the early radio %surveys did not have their redshifts measured. Hence the preferred values found %using these early radio-detected samples \citep{kar71,kar77}, would have been %determined by the radio survey alone, and not by a redshift selection effect. %The preferred redshifts (peaks) that are seen below z = 2 in the %\citet{kar71,kar77} sample are the same ones that are seen here. Many %investigators have previously sought to explain them using optical selection %effects \citep[and references therein]{bas05}, but have had only limited %success. [Roeder et al.]

Over the past 35 years many attempts have been made to attribute features in the redshift distribution to optical selection effects \citep{kar71,kar73,roe71,lak72,box84,kja78,bur01,baj03,baj04,bas05}. During this period the peak locations have changed above z = 2, so some of the previously claimed peak/selection effect correlations are no longer valid. Furthermore, observing methods have also changed (different filters used, etc.), but the peaks that were found to agree even with the redshift subcomponents predicted by eqn 1 remain \citep{bel02c,bel03b}. This is a strong argument that selection effects related to optical emission lines are not the cause of these peaks. Although \citet{roe71} claimed that broad peaks near z = 0.3 and z = 2 were due to the presence of strong lines in the observing window, when the number of available lines are plotted as a function of redshift the distribution produces only long period Fourier components. No power is seen at P = 0.7, although this could conceivably change with the addition of a few more sources at the appropriate locations, since the numbers involved were so small. This is not the case here, however, with 46,000 redshifts.

Also, \citet{tan05} report no evidence for a periodicity in the quasar redshift distribution obtained in the 2QZ survey. It is also readily apparent when the number of 2QZ quasars is plotted vs redshift \citep[see Fig 3]{cro04} that the peaks visible in the SDSS redshift distribution are not visible in the 2QZ distribution. We now have to ask why? When two independent surveys are obtained using optical identification techniques, if the peaks seen in one are due to optical selection effects, why do these selection effects not produce peaks in the other? Even if different filters are used the peaks should be present at some other redshift. This result suggests that optical selection effects are not the source of the peaks in the SDSS distribution. However, if the peaks are real, their absence in the 2QZ distribution must still be explained. \citet{tan05} explained this result by arguing that the 2QZ sample is more complete, and therefore free of selection effects. But this is a meaningless argument. In what sense is it more complete, especially below z = 2.4, and how does this completeness avoid the optical selection effects? Their argument is meaningful only if the valleys below z = 2.4 in the SDSS distribution are produced by the same type of selection effect as was suggested to explain the valleys at z = 2.7 and z = 3.5. However, this is unlikely, and to the best of our knowledge it has not been suggested in SDSS publications.

That there is little evidence for a period in the 2QZ redshift distribution can be explained if that survey found more sources with larger cosmological components present in their measured redshifts, filling in the valleys and smearing out the peaks. The aim of the 2QZ Survey was to measure redshifts for 25,000 optically-selected QSOs with b$_{J}$ $<20.85$ and z $<3$ \citep{cro04}. As noted above, the SDSS i-band limit was $<19.1$, and thus the 2QZ survey may have detected sources with larger cosmological components, although how big this effect might be is difficult to determine. This could have smeared out peaks in the 2QZ distribution, especially below z = 2.4, where the density of intrinsic components is higher. This is not an unrealistic conclusion since the broad baseline in Fig 6 is assumed to represent many sources in the SDSS sample that have cosmological, or Doppler ejection, components sufficiently large to smear out at least a portion of the peaks. The 2QZ survey was not designed to detect sources above z = 3 so it cannot be used to form any conclusions about the valleys that appear near z = 2.7 and 3.5 in the SDSS distribution.

%The suggestion that the Earth's atmosphere might affect the redshift %distribution \citep{roe71} also appears to be ruled out by the fact that the %redshift peaks are not seen in the 2QZ distribution.

\begin{figure}
\hspace{-1.0cm}
\vspace{-1.5cm}
\epsscale{1.0}
\plotone{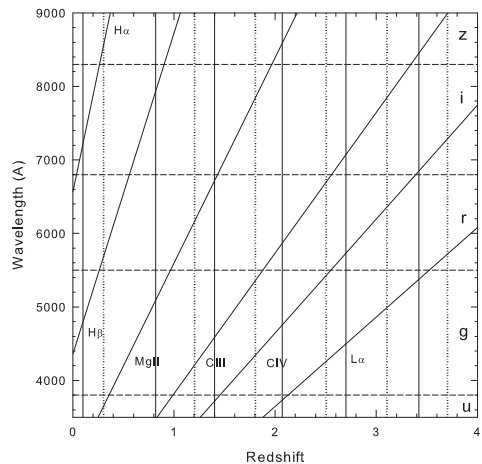}
%\plotone{filter2.eps}
\caption{\scriptsize{(solid oblique lines) Relation between redshift and observed wavelengths for strong emission lines. (dashed lines)Wavelenghts of gaps between SDSS filters. (solid vertical lines) Locations of valleys in the redshift distribution. (dotted lines) Locations of peaks in the distribution. The filter designation is given along the right edge.  \label{fig11}}}
\end{figure}

\begin{figure} 
\hspace{-1.0cm}
\vspace{-1.5cm}
\epsscale{1.0}
\plotone{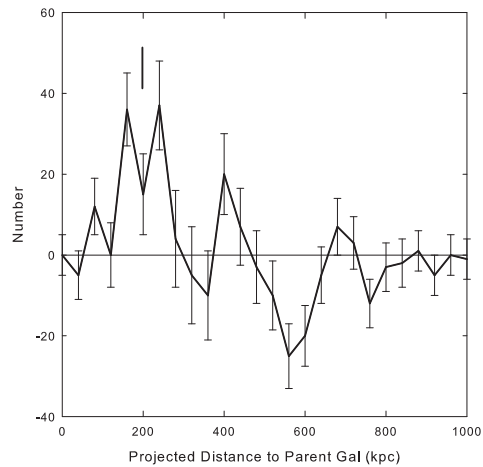}
%\plotone{tang2.eps}
\caption{\scriptsize{Difference between "true pairs" curve and randomly distributed galaxy curve taken from \citep[fig 7]{tan05}. The vertical bar indicates that there is a slight excess of sources near 200 kpc, the separation found by others to be typical for ejected sources. Error bars have been taken from \citet{tan05}. \label{fig12}}}
\end{figure}

\section{Discussion}

Recently \citet{tan05} used a quasar-galaxy pairing analysis to investigate the question of whether or not high-redshift quasars are likely to be born through ejection from a parent active galaxy. It is difficult to assess the significance of this approach in finding parent galaxies since the required assumptions are rather poorly known.
%First, these authors included only galaxies more distant than z = 0.01, and %there may now be good evidence to indicate that most of the quasars found in %early surveys may be closer than the distance implied by this redshift %\citep[see fig 3]{bel05}.
%Furthermore, the SDSS galaxy survey found galaxies with cosmological redshifts %mostly less than z$_{c}$ = 0.2. If, for example, ejected QSOs are intrinsically %at least 5 magnitudes less luminous than their parent galaxies, the SDSS quasar %search would be expected to find mainly objects closer than z$_{c}$ = 0.02 for %the same detection limit.
The \citet{tan05} analysis could thus have missed, or miss-identified, many of the parent galaxies, which could explain why the pairs they found differed little from what would be expected for a random distribution. In spite of this, although it was not pointed out by these authors, their pairs did show a slight excess near the expected value of 200 kpc (see their Fig 7). That result is reproduced here in Fig 12, which represents their \em true pair \em plot with their random galaxy curve subtracted from it. A typical projected separation value near 200 kpc was reported previously \citep{bur01,bur03}. If the objects are ejected uniformly in all directions, a higher number would be seen when they are ejected closer to the plane of the sky. Although \citet{tan05} concluded that QSOs are not ejected from active galaxies, it seems unlikely that the pair-finding technique they used could lead to a conclusion whose significance can approach that already obtained by others (Arp, the Burbidges, etc.), whose parent galaxy claims have been simultaneously backed up by other independent observations. As an example, we refer here to the case of the high-redshift QSO in front of the galaxy NGC 7319 \citep{gal04}. In fact, most of the conclusions reached by \citet{tan05} appear to have resulted because they have assumed that many of the values estimated in \citet{bel04} are much more accurate than they really are.

Here we have examined data samples containing a), the entire SDSS redshift distribution with 46,400 sources, b), the bottom half of that distribution containing approximately 40,000 sources and c), the upper half of that distribution containing approximately 6000 sources. All three showed evidence for the period predicted by eqn 1. It is also worth noting that a fourth source sample containing the 574 quasar redshifts used by \citet{kar71,kar77} was examined previously \citep{bel02c,bel03b} and it was found that the peaks in that distribution also correlated well with the preferred redshifts predicted by eqn 1. 

One of the most important aspects of the SDSS data has to be the huge number of sources involved. We conclude here that the significant power peak found for the full SDSS redshift distribution, with a period near $\Delta$z $\sim 0.7$, is real (i.e. not a statistical fluctuation), and is due either to selection effects in the data or to the presence of preferred redshifts. Because of the good agreement we find between the observed peaks and the preferred redshifts predicted by equation 1, it would seem likely that they have a common origin, and whatever that is, it must be able to explain both. This equation was derived empirically several years ago using a completely independent set of redshifts that were obtained by measuring the redshifts of X-ray excess QSOs near NGC 1068. The intrinsic redshift components were derived after removal of all ejection-related Doppler components, some of which produced redshift components as large a 0.366 \citep{bel02c}, that came from an ejection model \citep{bel02a,bel02b} that was based on the source positions on the sky relative to NGC 1068. \em As a result it is very unlikely that a common selection effect could have been involved. \em This may rule out selection effects as the common origin of the peaks in the SDSS redshift distribution and the preferred values predicted by eqn 1.

\section{Conclusions}

A power spectrum analysis of the distribution of over 46,000 SDSS quasar redshifts has been found to show a single, distinct power peak for redshift periods less than $\Delta$z = 1. The peak found corresponds to a redshift period of $\Delta$z $\sim$0.70. Not only is a distinct power peak observed, the locations of the peaks in the redshift distribution are in agreement with the preferred redshifts predicted by the intrinsic redshift equation (eqn 1). The power peak is detected in three different samples: the full SDSS sample, the lower half of the redshift distribution, and the upper half. We conclude that it is real, and is due either to the preferred redshifts predicted in the DIR model, or to selection effects. However, because of the way the intrinsic redshift relation was determined it seems unlikely that one selection effect could have been responsible for both.

\section{Acknowledgements} 

We thank an anonymous referee for several helpful suggestions.

Funding for the Sloan Digital Sky Survey (SDSS) has been provided by the Alfred P. Sloan Foundation, the Participating Institutions, the National Aeronautics and Space Administration, the National Science Foundation, the US Department of Energy, the Japanese Monbukagakusho, and the Max Planck Society. The SDSS website is http://www.sdss.org/. 

The SDSS is managed by the Astrophysical Research Consortium (ARC) for the Participating Institutions. The Participating Institutions are The University of Chicago, Fermilab, the Institute for Advanced Study, The Japan Participation Group, The Johns Hopkins University, the Korean Scientist Group (KSG), Los Alamos National Laboratory, the Max-Planck-Institute for Astronomy (MPIA), the Max-Planck-Institute for Astrophysics (MPA), New Mexico State University, University of Pittsburgh, University of Portsmouth, Princeton University, the United States Naval Observatory, and the University of Washington.

\clearpage

\begin{deluxetable}{cccccccc}
\tabletypesize{\scriptsize}
\tablecaption{Intrinsic Redshifts Predicted by Eqn 1 for z$_{iQ} < 4.5$. \label{tbl-1}}
\tablewidth{0pt}
\tablehead{
\colhead{($n$)} & \colhead{z$_{iQ}[N=1,n$]} & \colhead{z$_{iQ}[N=2,n$]} & \colhead{z$_{iQ}[N=3,n$]} &  \colhead{z$_{iQ}[N=4,n$]}
 & \colhead{z$_{iQ}[N=5,n$]} & \colhead{z$_{iQ}[N=6,n$]} & \colhead{z$_{iQ}[N=7,n$]}
}
\startdata

0 & 0.620 & 1.240 & 1.860 & 2.48 & 3.10 & 3.72 & 4.340 \\
1 & 0.558 & 1.178 & 1.798 & 2.418 & 3.038 & 3.658 & 4.278 \\
2 & 0.496 & 1.054 & 1.488 & 1.178 \\
3 & 0.434 & 0.868 & 0.558 \\
4 & 0.372 & 0.620 \\
5 & 0.310 & 0.310 \\
6 & 0.248 \\
7 & 0.186 \\

8 & 0.124 \\
9 & 0.062 \\

\enddata 
\end{deluxetable}

\clearpage

\end{document}